\documentclass[pre,showpacs,showkeys,preprintnumbers,amsmath,amssymb,superscriptaddress,twocolumn]{revtex4}
\pdfoutput=0 
\usepackage{graphicx}
\usepackage{amsfonts}
\usepackage{url}
\usepackage{epstopdf}
\usepackage{color}
\usepackage{bm}
\usepackage[colorlinks=true,linkcolor=blue,citecolor=red]{hyperref}%

\def\x{\bm{x}}
\def\v{\bm{v}}
\def\E{{\mathbb E}}

\def\Re{{\rm Re}}

\def\R{{\mathbb R}}

\definecolor{blue}{rgb}{0,0,1}

\begin{document}

\title{Statistical analysis of random trajectories of vibrated disks: \\
towards a macroscopic realization of Brownian motion}

\author{Yann Lanoisel\'ee}
  \email{yann.lanoiselee@polytechnique.edu}
\affiliation{Laboratoire de Physique de la Mati\`{e}re Condens\'{e}e (UMR 7643), \\ 
CNRS -- Ecole Polytechnique, University Paris-Saclay, 91128 Palaiseau, France}

\author{Guillaume Briand}
 \email{guillaume.briand@espci.fr}
\affiliation{Laboratoire Gulliver, UMR CNRS 7083, ESPCI Paris, PSL University, 10, rue Vauquelin, 75231 Paris cedex 05, France}

\author{Olivier Dauchot}
 \email{olivier.dauchot@espci.fr}
\affiliation{Laboratoire Gulliver, UMR CNRS 7083, ESPCI Paris, PSL University, 10, rue Vauquelin, 75231 Paris cedex 05, France}

\author{Denis~S.~Grebenkov}
 \email{denis.grebenkov@polytechnique.edu}
\affiliation{Laboratoire de Physique de la Mati\`{e}re Condens\'{e}e (UMR 7643), \\ 
CNRS -- Ecole Polytechnique, University Paris-Saclay, 91128 Palaiseau, France}

\affiliation{Interdisciplinary Scientific Center Poncelet (ISCP),%
\footnote{International Joint Research Unit -- UMI 2615 CNRS/ IUM/ IITP RAS/ Steklov MI RAS/ Skoltech/ HSE, Moscow, Russian Federation} \\
Bolshoy Vlasyevskiy Pereulok 11, 119002 Moscow, Russia}

\date{\today}

\begin{abstract}
We propose a macroscopic realization of planar Brownian motion by
vertically vibrated disks.  We perform a systematic statistical
analysis of many random trajectories of individual disks.  The
distribution of increments is shown to be almost Gaussian, with slight
deviations at large increments caused by inter-disk collisions.  The
velocity auto-correlation function takes both positive and negative
values at short lag times but rapidly vanishes.  We compare the
empirical and theoretical distributions of time averaged mean square
displacements and discuss distinctions between its mean and mode.
These well-controlled experimental data can serve for validating
statistical tools developed for the analysis of single-particle
trajectories in microbiology.
\end{abstract}




\maketitle

\section{Introduction}
One of the first reports of Brownian motion is attributed to the
Scottish botanist Robert Brown who observed in a microscope a
continuous jittery motion of minute particles ejected from the Clarkia
pollen grains suspended in water \cite{Brown1828}.  Since a more
systematic study by Jean Perrin \cite{Perrin1908,Perrin1909}, the
abundant experimental evidence of Brownian motion of microscopic
particles has been established
\cite{Frey05,Grebenkov07,Sackmann10,Brauchle}.  The mathematical
origin of this abundance lies in the central limit theorem which
implies a universal probabilistic description of motion at mesoscopic
time and length scales, regardless microscopic dynamics.  In turn,
experimental observations of Brownian motion in the macroscopic world
are rarer.  In fact, it is quite difficult to design an experiment
with macroscopic objects that would result in Brownian trajectories.
On one hand, the motion is strongly influenced by inertial effects,
resulting in ballistic segments of the trajectory at the macroscopic
scale (e.g., the motion of balls in a billiard).  On the other hand,
the number of interacting objects in a macroscopic system is much
smaller than the number of water molecules involved in the motion of a
microscopic particle, whereas the separation between the time scale of
an elementary displacement and the duration of the measurement is not
large enough.  As a consequence, the motion of macroscopic objects is
not enough randomized by their collisions.  In particular, the
dynamics of granular matter is typically far from Brownian motion
\cite{Jaeger96,Choi04,Grebenkov08,Chen09,Scalliet15}.  For instance,
there is a rather narrow range of packing fractions, for which the
motion of spherical beads is fluid-like: in the low density regime,
collisions between beads are rare while the mean free path is long so
that too large experimental setups would be needed to observe a
Brownian trajectory; in the high density regime, inter-bead collisions
are often but collective modes of motion (e.g., crystallization or
jamming) become dominant.

From the practical point of view, a well-controlled experimental
realization of a macroscopic diffusive motion with an excellent
statistics of long trajectories can serve as a benchmark for testing
various statistical tools developed for the analysis of single
particle trajectories (see
\cite{Qian91,Arcizet08,Metzler09,Michalet10,Berglund10,Michalet12,Vestergaard14,Gal13,Kepten15,Magdziarz11,Lanoiselee16,Lanoiselee17}
and references therein).  In fact, it is essential to disentangle
finite time average and finite sampling effects when performing single
probe experiments in biology (e.g., the intracellular transport or the
motion of proteins on cell membranes).  While statistical tools are
commonly tested on simulated trajectories, a macroscopic realization
of diffusive motions can present a rare opportunity to confront
simulations and theoretical results to an experimental situation with
true experimental noise, uncertainties, resolution issues, etc.

In this paper, we report an experimental observation of the diffusive
motion realized by macroscopic disks of 4 mm diameter on a vertically
vibrating plate (see Sec. \ref{sec:setup}).  Vibrations pump in the
system the kinetic energy that substitutes thermal energy that drives
the motion in a microscopic system.  We undertake a systematic
statistical analysis of the acquired trajectories of individual disks
(Sec. \ref{sec:analysis}).  In particular, we analyze the distribution
of one-step displacements, the ergodicity, the velocity
auto-correlation function, and the distribution of time averaged mean
square displacements (TAMSD).  This analysis shows that the
macroscopic motion of disks exhibits small deviations from Brownian
motion at short times but approaches it at longer times.  

\section{Experimental setup}
\label{sec:setup}

The experimental system, made of vibrated disks, has been described in
details previously \cite{Deseigne12}.  We recall here the key
ingredients of the set-up.  Experiments with shaken granular particles
are notoriously susceptible to systematic deviations from pure
vertical vibration and special care must be taken to avoid them.
First, to ensure the rigidity of the tray supporting the particles, we
use a $110$ mm thick truncated cone of expanded polystyrene sandwiched
between two nylon disks.  The top disk (diameter $425$ mm) is covered
by a glass plate on which lay the particles.  The bottom one (diameter
$100$ mm) is mounted on the slider of a stiff square air-bearing
(C40-03100-100254, IBSPE), which provides virtually friction-free
vertical motion and submicron amplitude residual horizontal motion.
The vertical alignment is controlled by set screws.  The vibration is
produced with an electromagnetic servo-controlled shaker
(V455/6-PA1000L,LDS), the accelerometer for the control being fixed at
the bottom of the top vibrating disk, embedded in the expanded
polystyren.  A $400$ mm long brass rod couples the air-bearing slider
and the shaker.  It is flexible enough to compensate for the alignment
mismatch, but stiff enough to ensure mechanical coupling.  The shaker
rests on a thick wooden plate ballasted with $460$ kg of lead bricks
and isolated from the ground by rubber mats (MUSTshock 100x100xEP5,
Musthane).  We have measured the mechanical response of the whole
setup and found no resonances in the window $70-130$ Hz.  We use a
sinusoidal vibration of frequency $f=95$ Hz and set the relative
acceleration to gravity $\Gamma = a (2\pi f)^2/g = 2.4$, where the
vibration amplitude $a$ at a peak acceleration is $100~\mu$m.  Using a
triaxial accelerometer (356B18, PCB Electronics), we checked that the
horizontal to vertical ratio is lower than $10^{-2}$ and that the
spatial homogeneity of the vibration is better than $1\%$.

The particles are micro-machined copper-beryllium disks (diameter $d =
4\pm 0.03$ mm).  The contact with the vibrating plate is that of an
extruded cylinder, resulting in a total height $h=2.0$ mm.  They are
sandwiched between two thick glass plates separated by a gap $H=2.4$
mm and laterally confined in an arena of diameter $320$ mm.  A CCD
camera with a spatial resolution of 1728 x 1728 pixels and standard
tracking software is used to capture the motion of the particles at a
frame rate of $25$ Hz.  In a typical experiment, the motion of the
disks is recorded during $600$ seconds, producing $15~000$ images.
The resolution on the position $\vec{r}$ of the particles is better
than $0.05$ particle diameter (i.e., $0.2$~mm).

In the following, particle trajectories are tracked within a circular
region of interest (ROI) of diameter $50d = 200$~mm far from the
border of the arena, where the long-time averaged density field is
homogeneous.  The average packing fractions $\phi$ measured inside the
ROI ranges from $0.3$ to $0.64$, and the total number of particles
ranges from $1000$ to $2500$.  As the onset of spatial order typically
takes place at $\phi_{\dagger} \simeq 0.71$, we always deal with a
liquid state.

\section{Statistical analysis}
\label{sec:analysis}

\begin{figure}
\begin{center}
\includegraphics[width=80mm]{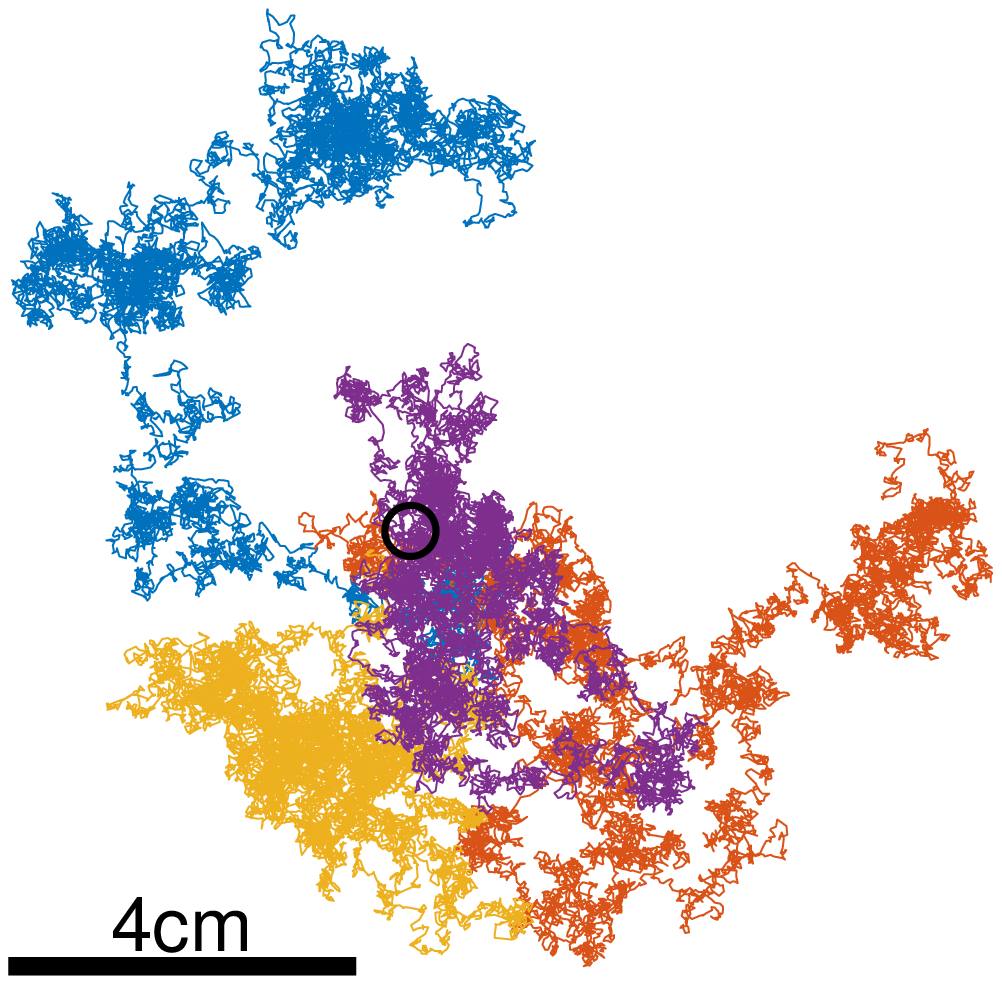} 
~ \vskip 5mm ~
\includegraphics[width=80mm]{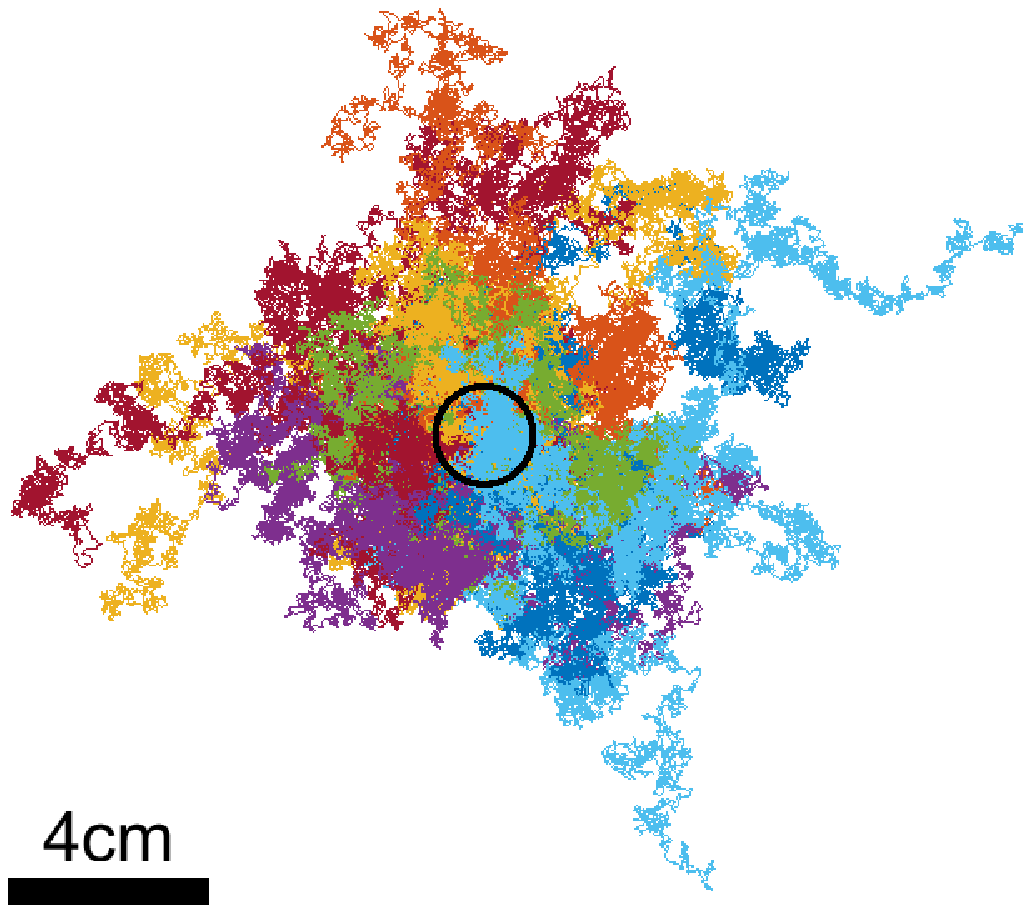} 
\end{center}
\caption{
(Color online) {\bf (Top)} Example of random trajectories of 4 disks
that were originally located close to each other (inside the black
circle) and then diffused during 600 seconds (i.e., 15~000 points in
each trajectory).  {\bf (Bottom)} Example of random trajectories of 34
disks that were originally located close to each other and then
diffused during 464.6 seconds (i.e., 11~615 points in each
trajectory).}
\label{fig:traj}
\end{figure}

We performed a systematic statistical analysis of the acquired random
trajectories.  Examples of such trajectories are shown in
Fig. \ref{fig:traj}.

\subsection{Data description}

We analyzed 14 samples with different surface packing fractions
$\phi$, ranging between $0.298$ and $0.637$ (Table \ref{tab:data}).
The time step (i.e., the duration of one displacement) is fixed by the
acquisition frequency: $\delta = 1/25~\textrm{Hz} = 0.04$~s.  The
positions are measured in units of the disk diameter, $d = 4$~mm.  To
avoid boundary effects, only the disks within the ROI were used for
the analysis.  In particular, a trajectory is terminated when the disk
leaves the ROI, and a new trajectory is initiated when a disk enters
the ROI.  As a consequence, the acquired trajectories have very
different lengths varying from $1$ to $15\,000$.  To improve the
statistical accuracy of our results, we discarded all the trajectories
whose length was shorter than $1000$.  The disks exhibited multiple
mutual collisions during the experiments.  Although the collective
motion of these disks might be studied as the dynamics of interacting
particles in a large phase space, we look at this problem from the
single-particle point of view and treat each disk as a single particle
interacting with its complex dynamic environment.  This view is
typical for single-particle tracking experiments in microbiology when
one can record only the motion of a labeled (e.g., fluorescent)
particle, whereas the dynamics of all other constitutes of the
cytoplasm remains inaccessible.

\begin{table}
\begin{center}
\begin{tabular}{|c |c | c | c |}  \hline
sample & $\phi$ & std/$d$ & $D$ (in mm$^2$/s) \\ \hline
1  & 0.298 & 0.0956 & 1.83 \\
2  & 0.324 & 0.0938 & 1.76 \\
3  & 0.350 & 0.0932 & 1.74 \\
4  & 0.376 & 0.1048 & 2.20 \\
5  & 0.402 & 0.1025 & 2.10 \\
6  & 0.428 & 0.1107 & 2.45 \\
7  & 0.454 & 0.1052 & 2.21 \\
8  & 0.480 & 0.1005 & 2.02 \\
9  & 0.507 & 0.0978 & 1.91 \\
10 & 0.533 & 0.0950 & 1.81 \\
11 & 0.559 & 0.0926 & 1.71 \\
12 & 0.585 & 0.0942 & 1.78 \\
13 & 0.611 & 0.0936 & 1.75 \\
14 & 0.637 & 0.0895 & 1.60 \\  \hline
\end{tabular}
\end{center}
\caption{
Summary of experimental data: the sample index, the surface fraction
$\phi$, the standard deviation of one-step one-dimensional increments
(in units of the disk diameter $d = 4$~mm), and the corresponding
diffusion coefficient: $D = {\rm std}^2/(2\delta)$, with $\delta =
0.04$~s.  For comparison the maximal disk packing fraction,
corresponding to the close-packed hexagonal lattice, is
$\pi/(2\sqrt{3}) \simeq 0.9069$; and the crystallization transition
for equilibrium hard disks takes place at $\phi_{\dagger}\simeq
0.71$. }
\label{tab:data}
\end{table}

\subsection{Distribution of increments}

\begin{figure}
\begin{center}
\includegraphics[width=85mm]{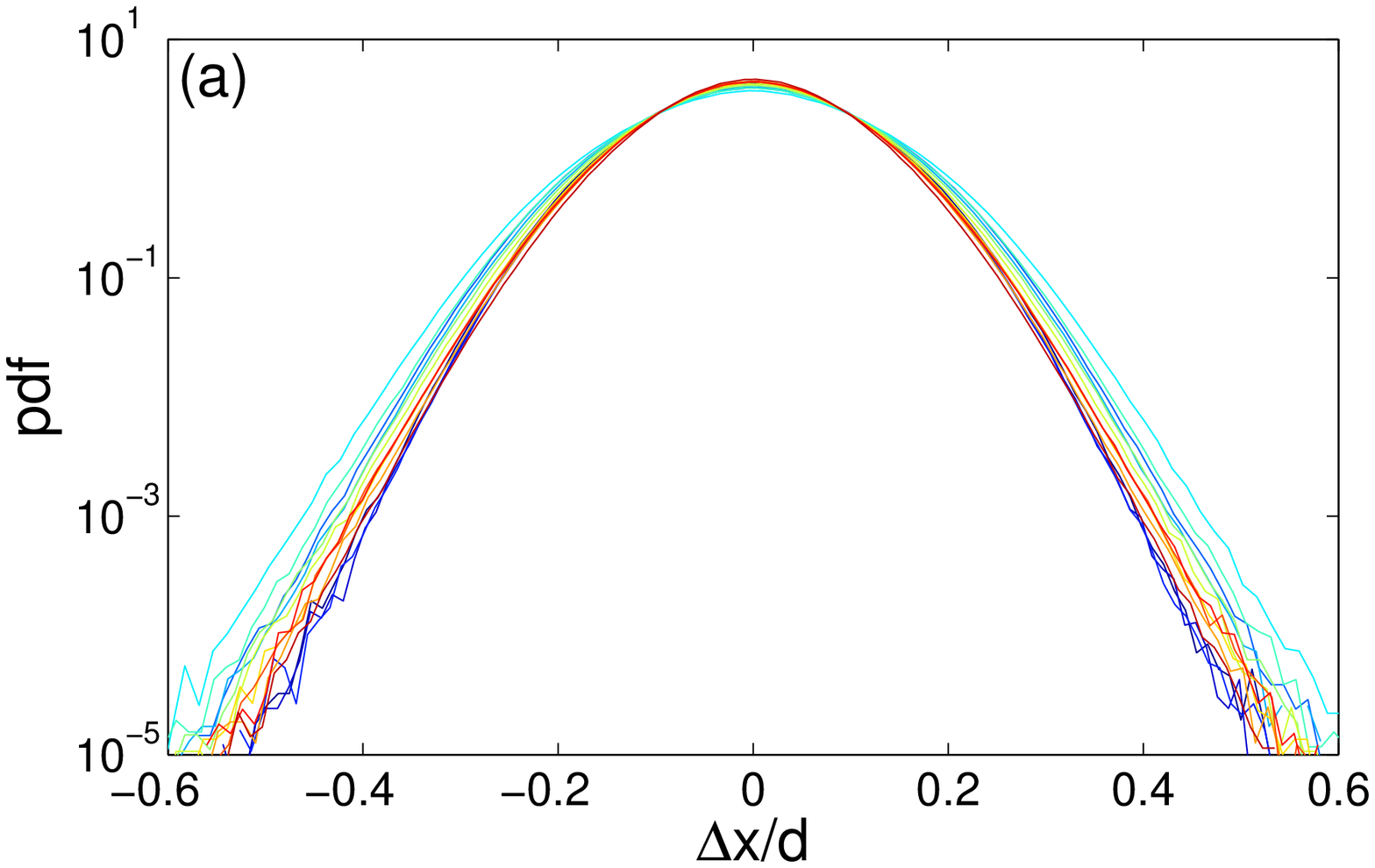} 
\includegraphics[width=85mm]{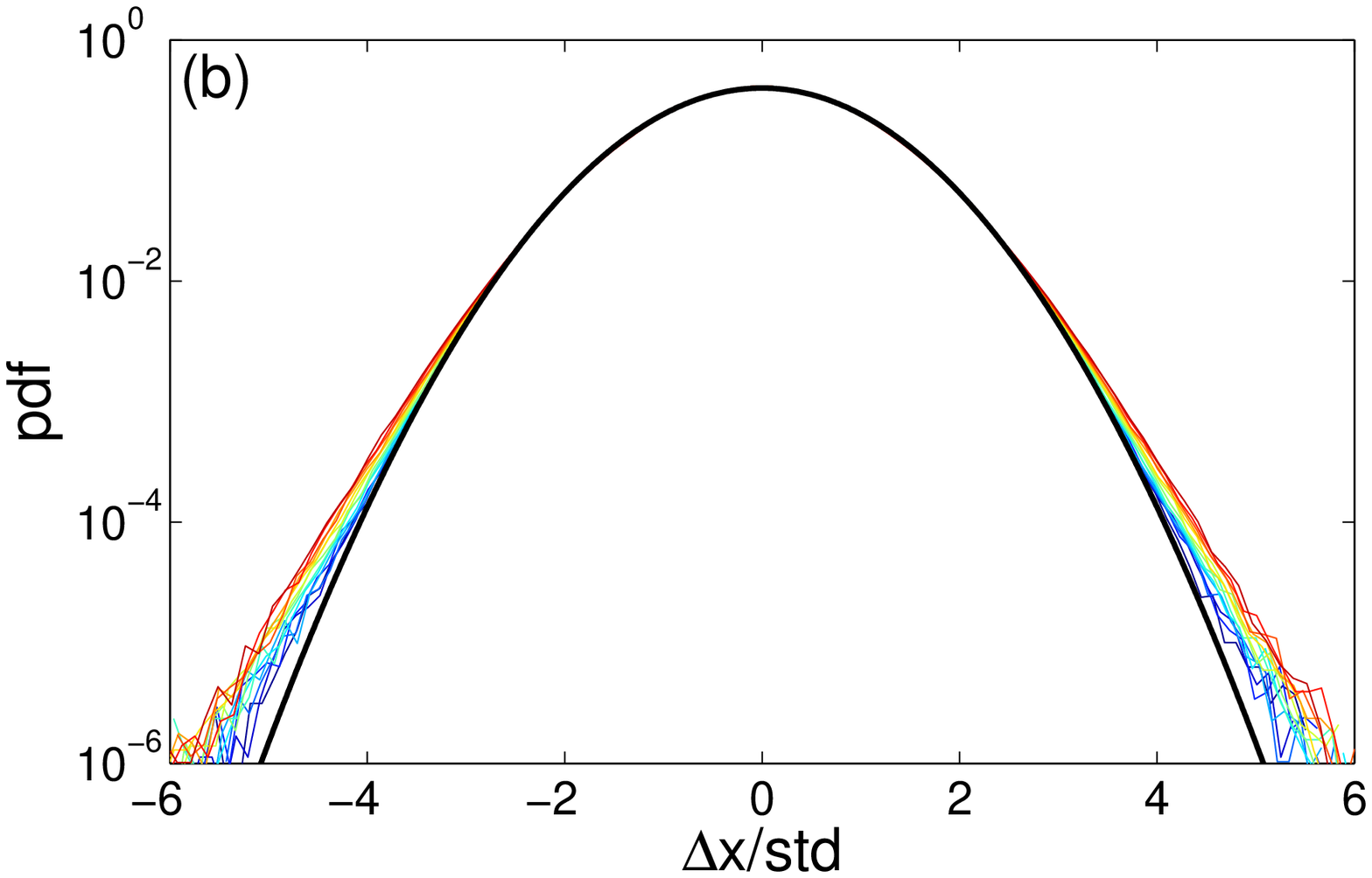} 
\end{center}
\caption{
(Color online) {\bf (a)} The empirical distributions of one-step
one-dimensional increments for 14 samples (in units of the disk
diameter, $d = 4$~mm).  {\bf (b)} The empirical distributions of
rescaled one-step increments for 14 samples.  Thick black curve shows
the standard Gaussian density $e^{-x^2/2}/\sqrt{2\pi}$.  Color of thin
curves changes from dark blue for the lowest packing fraction $\phi$
to dark red for the highest one.}
\label{fig:Gauss}
\end{figure}

We start by verifying whether the one-step increments obey a Gaussian
distribution.  For each sample, we collected the one-step increments
along $X$ and $Y$ axes for each trajectory in the sample and
constructed their histogram.  Having checked for the isotropy of the
statistics, we focus on one-dimensional increments and merge
increments along $X$ and $Y$ coordinates in order to get a
representative statistics even for large increments.  Figure
\ref{fig:Gauss}(a) shows these histograms (presented in the form of
probability densities at the semilogarithmic scale) for 14 samples.
These densities are close to each other and exhibit a parabolic shape
reminiscent of a Gaussian distribution.  The standard deviations of
one-step increments are summarized in Table \ref{tab:data}.  These
values are also close to each other and show no systematic dependence
on the packing fraction.  At first sight, there is no systematic
variation of probability densities with the packing fraction.  This
suggests that the randomness of motion essentially comes from the
rotational symmetry of the disk, which undergoes a displacement in a
random direction after each kick by the vibrating plate.  Note that
the frequency of plate vibrations is 4 times higher than the
acquisition frequency meaning that each displacement results from 4
random kicks.

Despite their delicate machining, the precise contact of the disks
with the vibrating plate is influenced by minor asperities, which
differ from disk to disk but also depend on the location of the disks
on the vibrating plate.  In order to reduce these factors of
diversity, we rescale the one-step increments from one trajectory by
the empirical standard deviation of these increments.  Such a
rescaling partly levels off eventual heterogeneities between
trajectories.  Once calculated, the rescaled increments along $X$ and
$Y$ coordinates are merged from different trajectories in each sample.
The obtained distributions are presented in Fig. \ref{fig:Gauss}(b).
One can see that the distributions for all 14 samples almost collapse
and remain close to the standard Gaussian density
$\exp(-x^2/2)/\sqrt{2\pi}$.  However, now that heterogeneities between
trajectories have been levelled off by the rescaling, one
distinguishes small but statistically significant deviations for large
increments.  These deviations progressively increase with the packing
fraction, and can therefore be attributed to disk-disk collisions.

\subsection{Ergodicity hypothesis}

\begin{figure}[b!]
\begin{center}
\includegraphics[width=85mm]{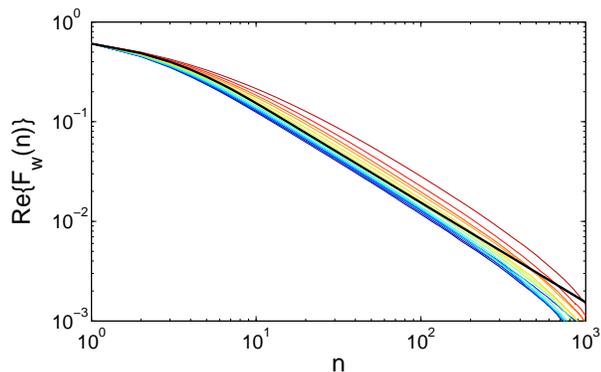} 
\end{center}
\caption{
(Color online) The real part of the ergodicity estimator,
$\Re\{\hat{F}_\omega(n)\}$ (with $\omega = 1/\sigma$), averaged over
trajectories in each of 14 samples (thin lines).  Color of thin curves
changes from dark blue for the lowest packing fraction $\phi$ to dark
red for the highest one.  Thick black line shows the mean value of
this estimator for Brownian motion.  }
\label{fig:ergo}
\end{figure}

We analyze whether the system of vibrated disks can be considered as
being at equilibrium.  In practice, we test the ergodicity hypothesis
which is a necessary but not sufficient condition for equilibrium.
The ergodicity hypothesis claims that the ensemble average over many
particles is equal to the time average over an (infinitely) long
trajectory of one particle.  Under the stationarity hypothesis of the
motion, we employ the ergodicity estimator $\hat{F}_\omega(n)$
\cite{Magdziarz11,Lanoiselee16}
\begin{equation}   
\label{eq:F}
\hat{F}_\omega(n) \equiv \frac{1}{n} \sum\limits_{k=1}^{n} \hat{E}_\omega(k),
\end{equation}
with
\begin{eqnarray}
\nonumber
\hat{E}_\omega(n) & \equiv & \frac{1}{N-n+1} \sum\limits_{k=0}^{N-n} e^{i\omega [X(k+n) - X(k)]} \\
\label{eq:E}
&-& \frac{1}{N(N+1)} \left|\sum\limits_{k=0}^{N} e^{i\omega [X(k)-X(0)]} \right|^2 + \frac{1}{N} .
\end{eqnarray}
where $X(k)$ are successive $X$ coordinates of the points $\x_k$ along
a given trajectory of length $N$ (the same analysis was performed for
the $Y$ coordinate, not shown).  The first term can be interpreted as
the time averaged characteristic function of the increment
$X(k+n)-X(k)$ at lag time $n$, while the second term ensures that the
estimator is strictly $0$ for a constant process $X(n) = X_0$ (in
addition, the mean estimator is strictly $0$ for a process with
independent $X(n)$).  For Brownian motion, the mean value of the
estimator is \cite{Lanoiselee16}:
\begin{equation}
\E \{ \hat{F}_\omega(n)\} = q \frac{1 - q^n}{n(1-q)} + O(1/N) ,
\end{equation}
where $q = e^{-\omega^2\sigma^2/2}$ and $\sigma^2$ is the variance of
one-step increments.  To eliminate the effect of length scale, we set
$\omega = 1/\sigma$ that is equivalent to rescaling the trajectory by
the standard deviation $\sigma$.

Figure \ref{fig:ergo} shows the real part of the ergodicity estimator
averaged over all the trajectories in each of 14 samples.  For small
$n$, higher the packing fraction, slower the decrease of the estimator
with $n$.  However, for large $n$ the $1/n$ scaling predicted in the
case of the Brownian motion is recovered and we can safely formulate
the hypothesis that ergodicity is satisfied.

\subsection{Velocity auto-correlations}
\label{sec:VACF}

We also study the velocity auto-correlations function (VACF) which is
defined as
\begin{equation}
C(t) = \langle \v(t) \cdot \v(0) \rangle \,,
\end{equation}
where $\v(t)$ is the velocity at time $t$, and $\langle \cdots
\rangle$ is the ensemble average.  In the experimental setting, the
positions are recorded with the time step $\delta = 0.04$~s, so that
$t = n \delta$, and the velocity is proportional to the one-step
increment: $\v(n\delta) = (\x_{n+1} - \x_n)/\delta$, with $\x_n =
\x(n\delta)$.  To improve statistics, we combine the time average
along the trajectory of each disk and the ensemble average over many
trajectories:
\begin{equation}
C(n\delta) = \frac{1}{M\delta^2}\sum\limits_{m=1}^M  \frac{1}{N_m-n-1} \hspace*{-1mm} \sum\limits_{k=1}^{N_m-n-1} 
\bigl(\Delta\x_{n+k}^{(m)} \cdot \Delta\x_k^{(m)}\bigr) ,
\end{equation}
where $\Delta\x_n^{(m)} = \x^{(m)}_{n+1} - \x^{(m)}_n$ is the $n$-th
one-step increment of the $m$-th disk, $M$ is the number of disks in a
sample, and $N_m$ is the length of the $m$-th trajectory.

\begin{figure}
\begin{center}
\includegraphics[width=85mm]{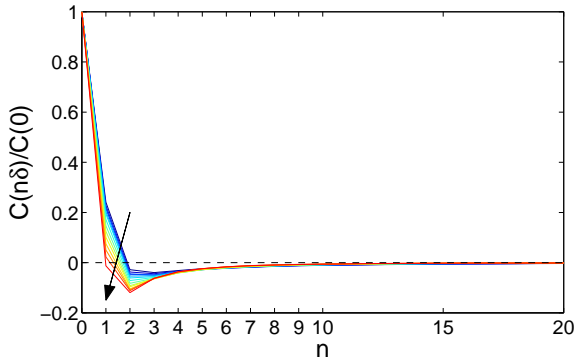} 
\end{center}
\caption{
The normalized VACF, $C(n\delta)/C(0)$, as a function of the lag time
$n\delta$, for 14 samples.  Color changes from dark blue for the
lowest packing fraction $\phi$ to dark red for the highest one.  This
change is also indicated by an arrow. }
\label{fig:VACF}
\end{figure}

Figure \ref{fig:VACF} shows the normalized VACF, $C(n\delta)/C(0)$,
which varies between $-1$ and $1$, as a function of the lag time
$n\delta$.  For all considered samples, the VACF rapidly decreases
with time and becomes close to zero for $n \gtrsim 10$.  By
construction, the normalized VACF is equal to $1$ at $n=0$.  Positive
auto-correlations at lag time $n=1$ can potentially be attributed to
inertial effects.  The negative auto-correlations observed for $n > 1$
take their root in an excess of reverse bouncing of the disks when
they successively hit the trail, but not only.  Since they become more
pronounced when the packing fraction increases, they should also come
from collisions.  In all cases, although the successive increments
exhibit small but noticeable correlations, they drop very rapidly as
the lag time increases.  We recall that the normalized VACF for a
discrete-time Brownian motion (a random walk) is $1$ for $n = 0$ and
$0$ otherwise.  Strictly speaking, the disk trajectories acquired at
time step $\delta = 0.04$~s are therefore not Brownian but remain
close to Brownian ones.

\subsection{Estimation of diffusion coefficient}

Now we focus on the time averaged mean square displacement, which is
the most common statistical tool to probe diffusive properties of
single-particle trajectories \cite{Gal13}.  The TAMSD with the lag
time $n$ over a trajectory of length $N$ is defined as
\begin{equation}
\chi_{n,N} = \frac{1}{N-n} \sum\limits_{k=1}^{N-n} \|\x_{k+n} - \x_{k}\|^2 .
\end{equation}
If $\x_k$ are positions of planar Brownian motion with diffusion
coefficient $D$, then the ergodicity of this process implies that
\begin{equation}
\chi_{n,N} \xrightarrow[N \to \infty]{}  4Dt = 4D n\delta ,
\end{equation}
whereas the variance of $\chi_{n,N}$ vanishes as $N\to\infty$
\cite{Qian91,Grebenkov11}.  In other words, the TAMSD allows one to
estimate the diffusion coefficient $D$ from a {\it single} random
trajectory, and longer the trajectory, better the estimation.

\begin{figure}
\begin{center}
\includegraphics[width=85mm]{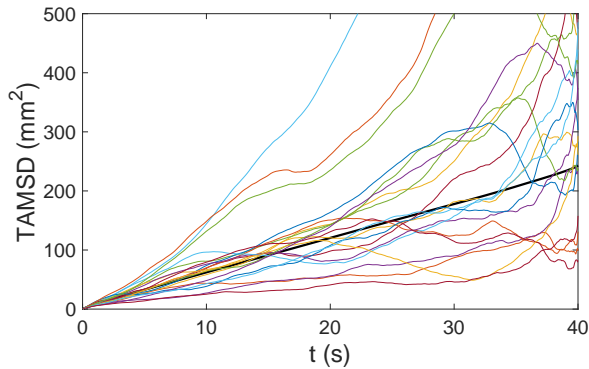} 
\end{center}
\caption{
The time averaged MSD, $\chi_{n,N}$, as a function of the lag time $t
= n\delta$ (with $n$ varying from $1$ to $1000$), for 10 trajectories
of length $N = 1001$ from the sample 1 with the lowest packing
fraction.  Black thick line shows the ensemble average of TAMSD over
12444 trajectories of length $N = 1001$ in this sample.  The fitted
diffusion coefficient from this line is $1.47$~mm$^2$/s. }
\label{fig:TAMSD}
\end{figure}

For a fixed $N$, the smallest variance (and thus the best estimation)
corresponds to $n = 1$, in which case $\chi_{1,N}$ is the estimator of
the variance of increments.  This estimator is known to be optimal for
the case of Brownian motion, i.e., it is the best possible way to
estimate the diffusion coefficient
\cite{Cramer,Grebenkov11,Grebenkov13}.  In practice, however, even if
the studied particle is supposed to undergo Brownian motion, the
acquired trajectory can be altered by various ``measurement noises''
such as localization error, electronic noise, drift or vibrations of
the sample, post-processing errors, etc.  When some of these noises
are anticipated, the estimator can be adapted to provide the (nearly)
optimal estimation 
\cite{Michalet10,Berglund10,Michalet12,Grebenkov13,Vestergaard14}.
However, the Brownian character of the studied but yet unknown process
is not granted and has to be checked from the analysis of the TAMSD.
In this situation, the rule of thumb consists in plotting the TAMSD
versus the lag time $n$ to first check the linear dependence and then
to estimate the diffusion coefficient from the slope of the linear
plot.  Given the randomness of the TAMSD, this procedure can bring
biases and additional statistical errors.  Moreover, since
fluctuations of the TAMSD grow with $n$ (see
\cite{Qian91,Grebenkov11}), the fit is often limited to small $n$.
Figure \ref{fig:TAMSD} illustrates large fluctuations of the TAMSD
estimator around the ensemble averaged TAMSD which exhibits a linear
growth with $n$.  As a consequence, an accurate estimation of the
diffusion coefficient from a {\it single} trajectory is only possible
over a narrow range of small lag times $n$.  Note that the diffusion
coefficient fitted by the ensemble average, $1.47$~mm$^2$/s, is
smaller than that estimated from the standard deviation of one-step
increments, $1.83$~mm$^2$/s (see Table \ref{tab:data}).  This
discrepancy can be caused by eventual noises (that would affect the
standard deviation of one-step increments) and auto-correlations (that
would affect the TAMSD).

\subsection{Distribution of TAMSD}

One of the significant advantages of single-particle tracking is the
possibility to infer information from single events, without ensemble
averages.  This is particularly important in microbiology because many
events in a cell life are triggered by a small number of molecules.
Even when many particles are tracked simultaneously, they explore
different spatial regions of the cell and experience different
interactions with the intracellular environment.  If infered properly,
such heterogeneities may bring a much more detailed information about
the cell than an ensemble average.  The estimation of the diffusion
coefficient from each single trajectory naturally leads to their
distribution \cite{Duits09,Otten12,Nandi12}.  However, it is important
to stress that the experimentally obtained distributions include two
sources of randomness: (i) the biological variability and (ii) the
intrinsic randomness of the TAMSD estimator obtained from a single
finite length trajectory.  As a consequence, a proper biological
interpretation of such distributions requires to disentangle two
sources and, ideally, to remove the second one.  This correction needs
the knowledge of the distribution of the TAMSD estimator.

The distribution of TAMSD in the biological context was first studied
via numerical simulations by Saxton \cite{Saxton93,Saxton97}.  A more
general theoretical analysis of TAMSD for Gaussian processes was later
performed in Refs. \cite{Grebenkov11,Andreanov12,Grebenkov13,Sikora17}.  
We compute the distribution numerically via the inverse Fourier
transform of the characteristic function of TAMSD for which the exact
matrix formula was provided in Ref. \cite{Grebenkov11}.  This
computation was shown to be fast and very accurate.

The theoretical distribution of TAMSD for Brownian motion can be
compared to the empirical distribution of TAMSD obtained from the
trajectories of disks.  On one hand, this comparison allows one to
check to which extent the acquired trajectories are close to Brownian
motion.  On the other hand, one can investigate in a well-controlled
way the applicability of the theoretical distribution to experimental
data.

\begin{figure*}
\begin{center}
\includegraphics[width=85mm]{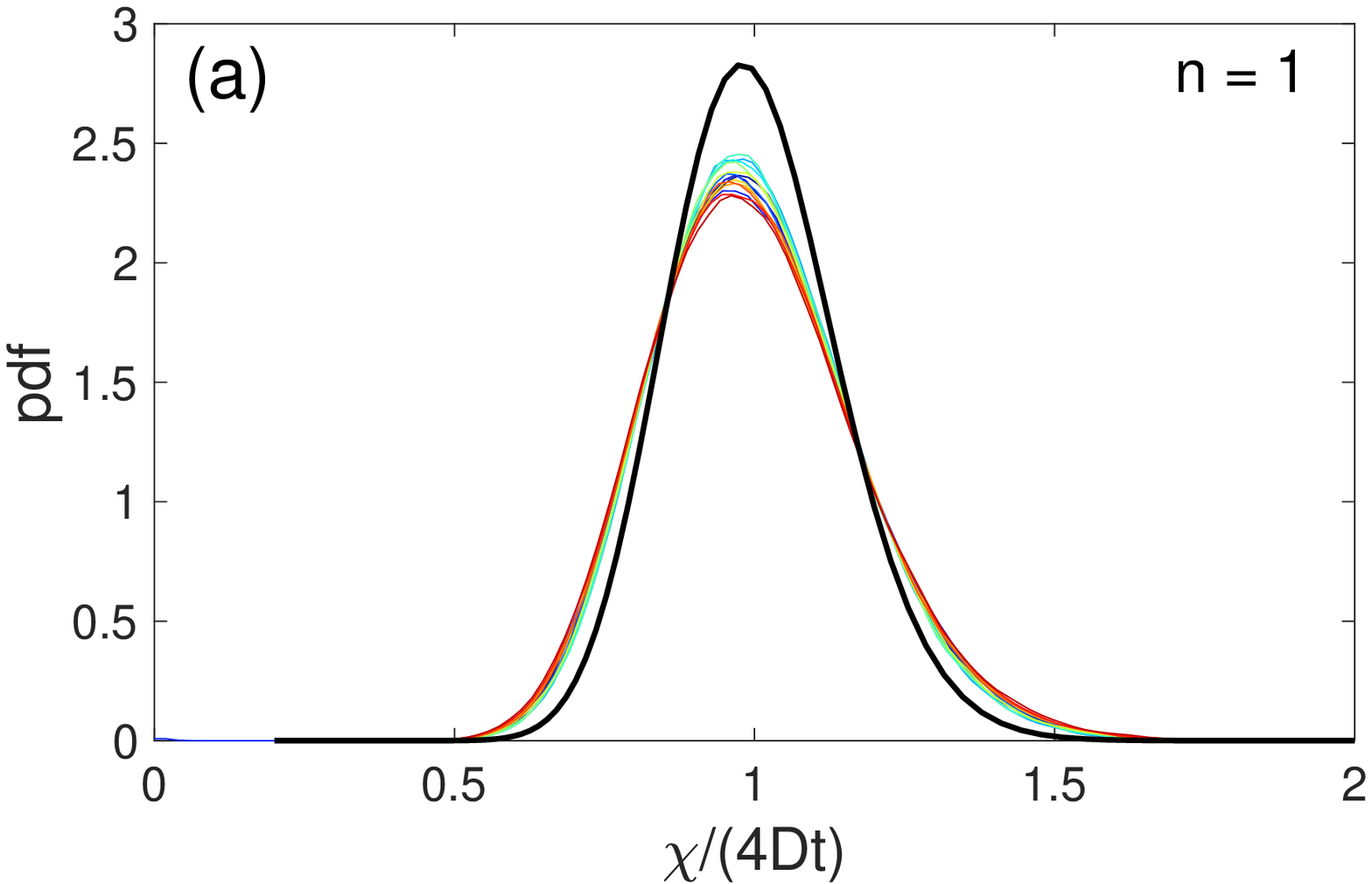} 
\includegraphics[width=85mm]{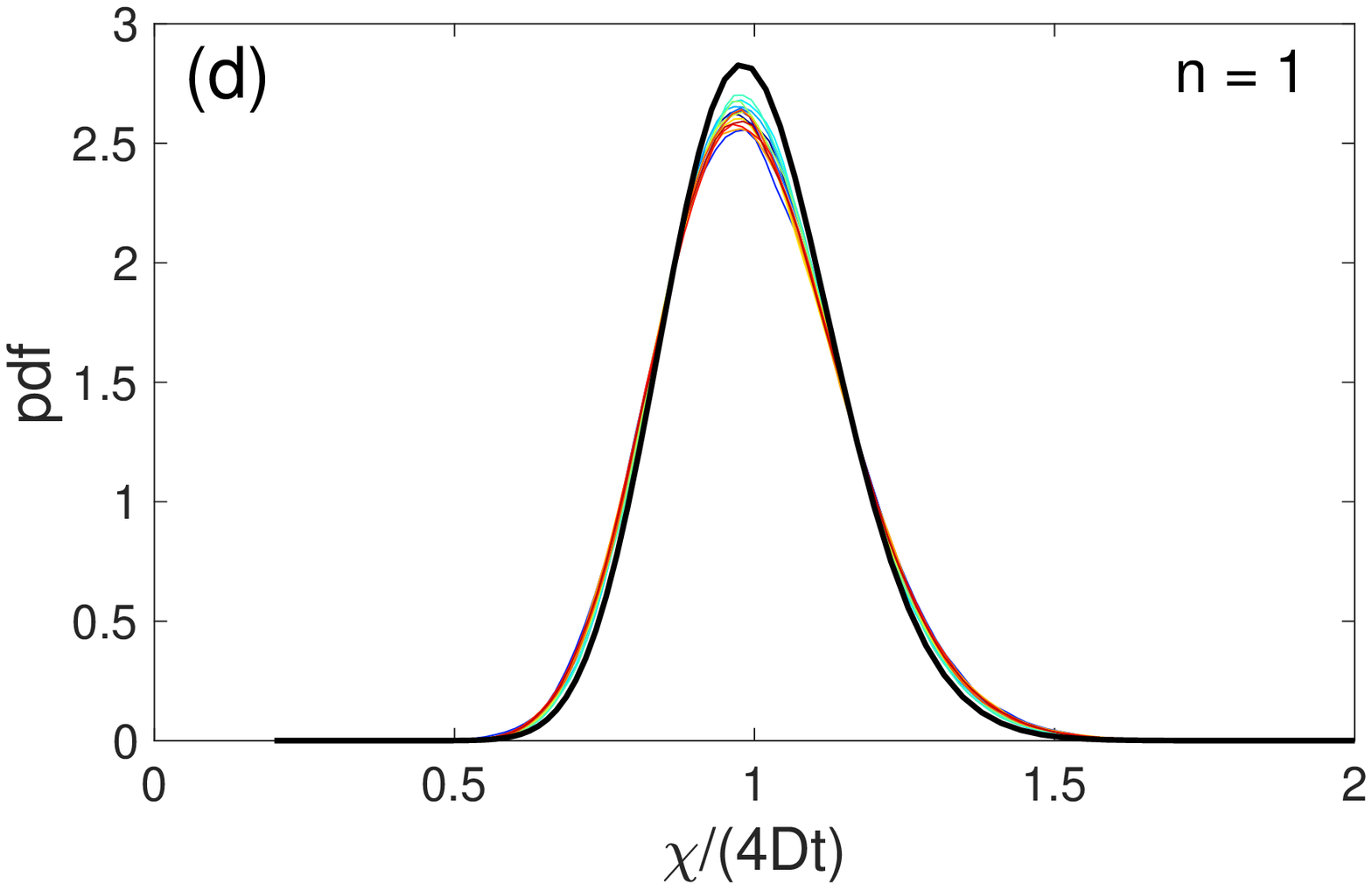} 
\includegraphics[width=85mm]{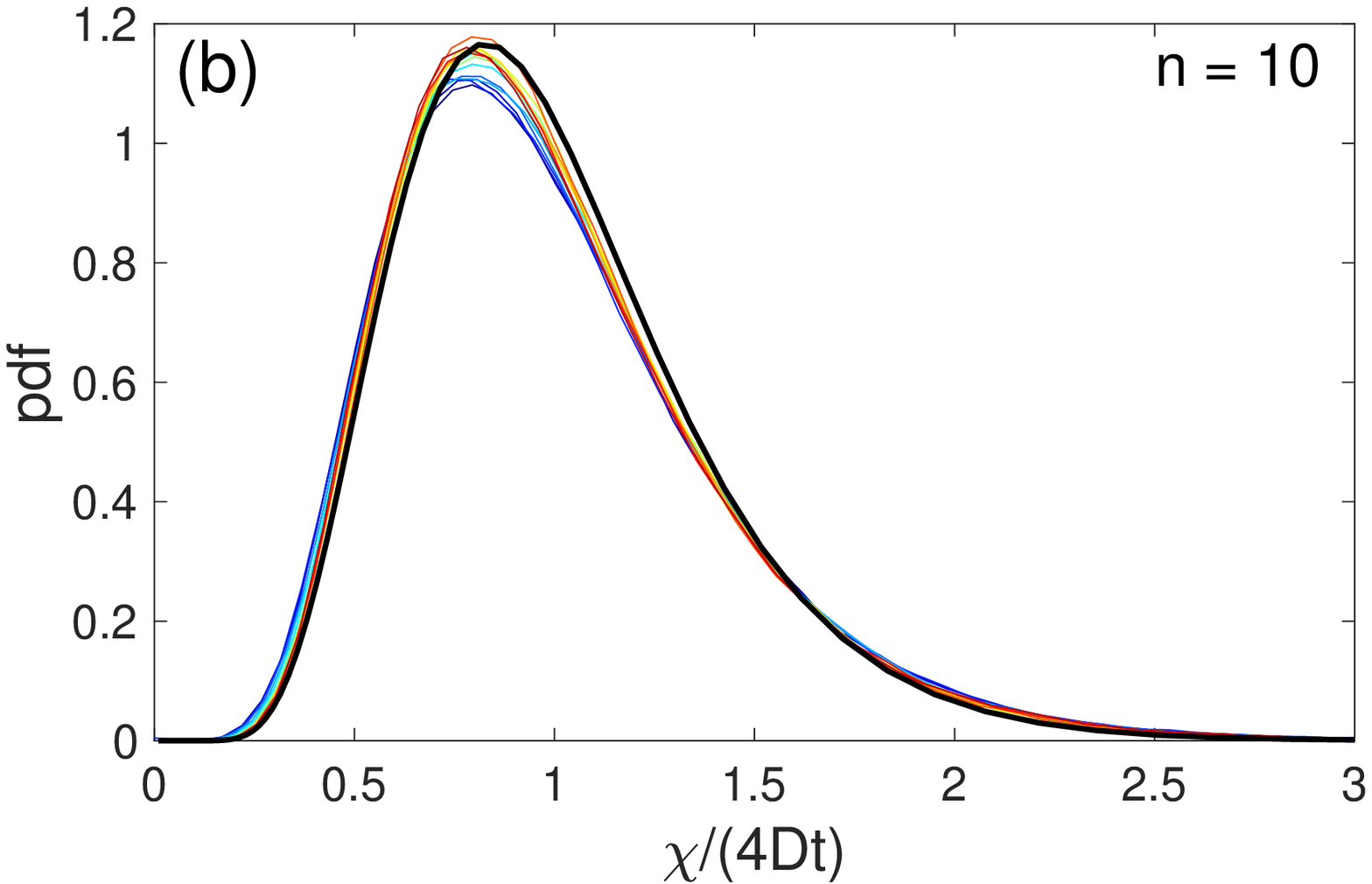} 
\includegraphics[width=85mm]{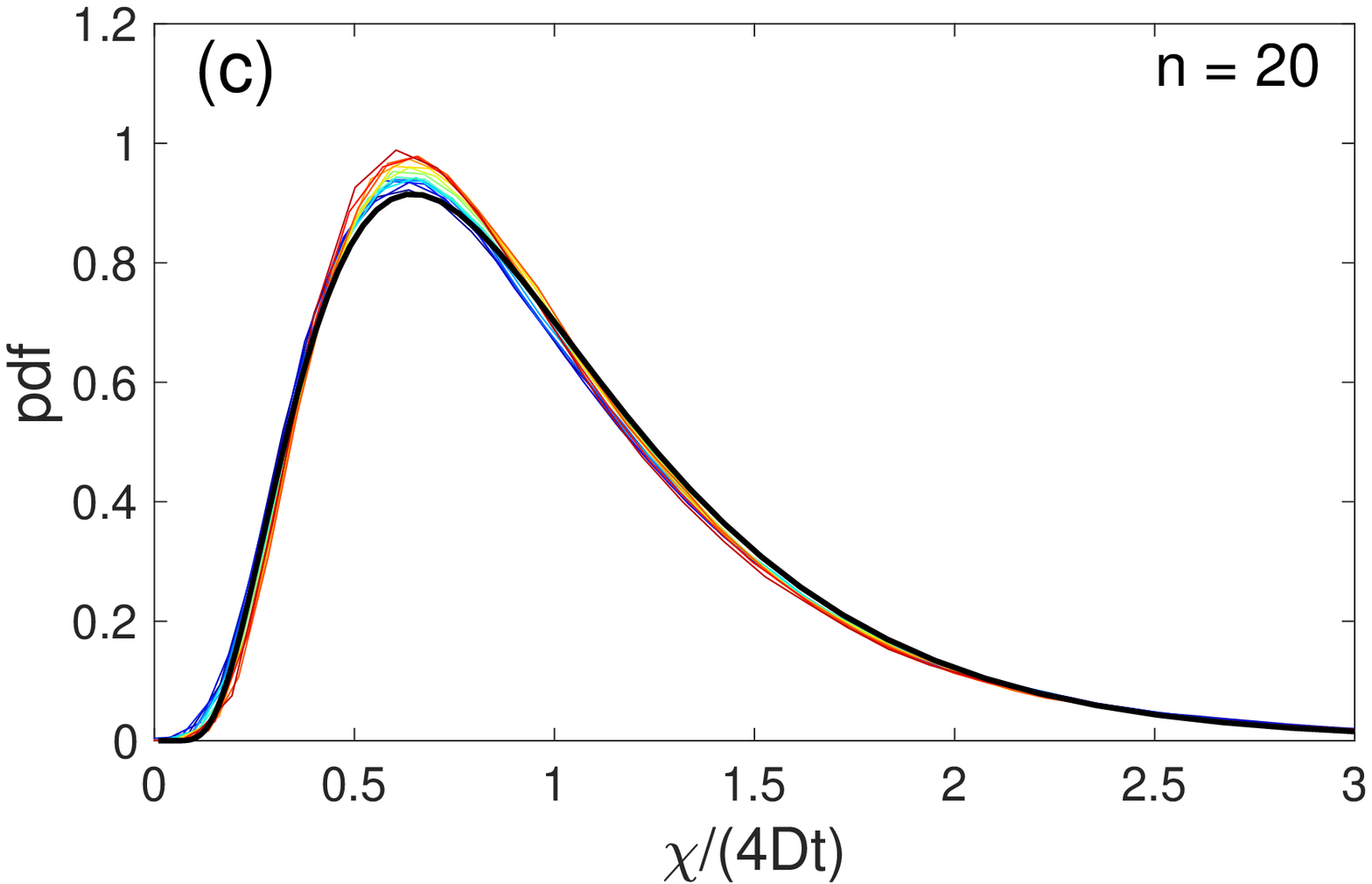} 
\end{center}
\caption{
(Color online) {\bf (a,b,c)} Probability density functions of the
rescaled TAMSD, $\chi_{n,N}/(4Dt)$, with $N = 100$, $t = n\delta$, and
$n = 1$ (a), $n = 10$ (b), and $n = 20$ (c), for 14 samples (thin
lines), and the theoretical ones for Brownian motion (thick black
line).  Color of thin curves changes from dark blue for the lowest
packing fraction $\phi$ to dark red for the highest one. {\bf (d)}
Probability density functions of the rescaled TAMSD with $N = 100$, $t
= n\delta$, and $n = 1$, for 14 reshuffled samples. }
\label{fig:Ddist_all}
\end{figure*}

Figure \ref{fig:Ddist_all}(a) shows the empirical distribution of
TAMSD with the lag time $n = 1$ obtained by splitting each trajectory
into fragments of length $N = 100$.  This artificial splitting is
performed to be closer to the common situation in biological
applications, when the acquired trajectories are rather short.
Moreover, such splitting significantly improves the statistics of the
TAMSD.  We compare the probability density functions of TAMSD among 14
samples and with the theoretical curves for Brownian motion.  One can
see notable deviations from the theoretical distribution, indicating
that the acquired trajectories are not Brownian, in agreement with the
analysis of Sec. \ref{sec:VACF}.  The two plausible reasons for the
observed deviations are: (i) auto-correlations of increments at small
lag times (as seen in Fig. \ref{fig:VACF}), and (ii) small deviations
from the Gaussian distribution of increments (as seen in
Fig. \ref{fig:Gauss}).  To check for the first reason, we plot in
Fig. \ref{fig:Ddist_all}(b,c) the distributions of the TAMSD with
larger lag times $n = 10$ and $n = 20$, at which the VACF was
negligible.  One gets thus a much better agreement with the
theoretical distribution.

In order to check the second reason of deviations (weak
non-Gaussianity), the increments of all trajectories in each sample
were randomly reshuffled to fully destroy auto-correlations, and then
new artificial trajectories were constructed from these increments.
If the original increments were correlated Gaussian variables with the
same variance, such a procedure would yield independent identically
distributed Gaussian variables so that the resulting trajectories
would represent Brownian motion.  In this case, a perfect agreement
between empirical and theoretical curves would be expected.  Figure
\ref{fig:Ddist_all}(d) shows empirical and theoretical
distributions of TAMSD at the lag time $n = 1$ for reshuffled samples.
The agreement is not perfect but is much better than in
Fig. \ref{fig:Ddist_all}(a).  Small residual deviations can
potentially be attributed to weak non-Gaussianity of the distribution
of increments.

\subsection{Mean versus the most probable TAMSD}

The nonsymmetric shape of the distribution of TAMSD implies that the
mean value of the TAMSD is different from its mode, i.e., the most
probable value or, equivalently, the position of the maximum of the
PDF.  This difference becomes particularly important for the analysis
of single particle trajectories.  When the sample of such individual
trajectories is large, the empirical mean of TAMSD estimated from
these trajectories is close to the expectation.  In turn, when the
TAMSD is estimated from few trajectories (or even from a single
trajectory), it is more probable to observe a random realization near
the maximum of the PDF.  This issue, which was not relevant for
symmetric distribution (e.g., a Gaussian distribution), may become an
important bias in the analysis of TAMSD.

As discussed in Ref. \cite{Grebenkov11}, the distribution of TAMSD for
Brownian motion is wider and more skewed for larger $n/N$.  Moreover,
the difference between the mean and the mode also grows with $n/N$.
As suggested in \cite{Grebenkov11}, the distribution of TAMSD for
Brownian motion and some other centered Gaussian processes (like
fractional Brownian motion) can be accurately approximated by a
generalized Gamma distribution, which has a simple explicit PDF
\begin{equation}  \label{eq:gamma}
p(z) = \frac{z^{\nu-1} \exp(-a/z - z/b)}{2(ab)^{\nu/2} K_{\nu}(2\sqrt{a/b})}  \quad (z > 0),
\end{equation}
with three parameters: $a \geq 0$, $b > 0$, and $\nu \in \R$ (here
$K_\nu(z)$ is the modified Bessel function of the second kind).  The
moments of this distribution can be expressed as
\begin{equation}  \label{eq:moments}
\langle [\chi_{n,N}]^k \rangle = (ab)^{k/2} \frac{K_{\nu+k}(2\sqrt{a/b})}{K_{\nu}(2\sqrt{a/b})}  \quad (k=1,2,3,\ldots),
\end{equation}
whereas the mode is
\begin{equation}
\chi_{n,N}^{\rm mode} = \frac{\sqrt{(1-\nu)^2 b^2 + 4ab} - (1-\nu)b}{2} \,.
\end{equation}
For a given empirical distribution of TAMSD, the first three moments,
evaluated directly from the data, can be used to calculate the
parameters $a$, $b$ and $\nu$ by solving numerically the system of
three equations in Eqs. (\ref{eq:moments}) for $k=1,2,3$.  In other
words, one does not need to fit the empirical distribution in order to
get this approximation.

\begin{figure}
\begin{center}
\includegraphics[width=85mm]{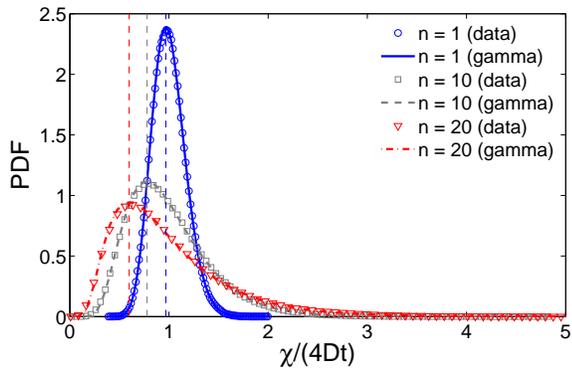}  
\end{center}
\caption{
(Color online).  Probability density functions of the rescaled TAMSD,
$\chi_{n,N}/(4Dt)$, for the sample with the lowest packing fraction,
with $N = 100$, $t = n\delta$, and $n = 1$ (blue circle), $n = 10$
(gray squares), and $n = 20$ (red triangles).  Lines show the
generalized gamma distribution $p(z)$ from Eq. (\ref{eq:gamma}) whose
parameters were obtained from the first three moments (see the text):
$a = 3.27$, $b = 0.03$, $\nu = 28.13$ ($n = 1$), $a = 1.94$, $b =
0.27$, $\nu = 1.34$ ($n = 10$), and $a = 1.03$, $b = 0.58$, $\nu =
0.31$ ($n = 20$).  The modes of these distributions are shown by
vertical dashed lines: $0.97$ ($n = 1$), $0.78$ ($n = 10$), and $0.60$
($n = 20$), whereas the mean is fixed to be $1$ by rescaling. }
\label{fig:pdf_gamma}
\end{figure}

Figure \ref{fig:pdf_gamma} shows the pdf of the TAMSD for the
trajectories with the lowest packing fraction, with the sample length
$N = 100$ and three lag times, $n = 1$, $n = 10$ and $n = 20$ (shown
by symbols).  From these empirical data, we evaluated the first three
moments and calculated the parameters $a$, $b$ and $\nu$ of the
generalized gamma distribution (shown by lines).  The excellent
agreement validates the use of this theoretical approximation even for
experimental trajectories.  

\section{Conclusion}

We proposed a macroscopic realization of planar Brownian motion by
vertically vibrated disks.  We performed a systematic statistical
analysis of many random trajectories of individual disks.  The
distribution of one-step increments was shown to be almost Gaussian.
Since small deviations at large increments increase with the disk
packing fraction, they were attributed to inter-disk collisions.  The
velocity auto-correlation function was positive at the lag time $n =
1$ and took negative values at $n > 1$ that rapidly vanish with $n$.
We also analyzed the behavior of the time averaged mean square
displacement as a function of the lag time, and its fluctuations from
one trajectory to another.  We compared the empirical and theoretical
distributions of TAMSD and revealed the sensitivity of this
distribution at small lag times to eventual auto-correlations and weak
non-Gaussianity.  We also verified that the empirical distribution can
be accurately approximated by the generalized Gamma distribution.
Finally, we discussed distinctions between the mean TAMSD and the mode
of its distribution.  These well-controlled experimental data can
serve for validating statistical tools developed for the analysis of
single-particle trajectories in microbiology.

\begin{acknowledgments}
DG acknowledges the support under Grant No. ANR-13-JSV5-0006-01 of the
French National Research Agency.
\end{acknowledgments}

\end{document}